# The geopolitics behind the routes data travels: a case study of Iran

*Loqman Salamatian, Frederick Douzet, Kevin Limonier, Kavé Salamatian*

## I ) Introduction:

The global expansion of the Internet has brought many challenges to the field of geopolitics. According to the definition of Yves Lacoste, geopolitics studies rivalries of power and influence over a territory at various levels of analysis [Lacoste 1993]. It analyses the dynamics of a conflict over a territory, the contradictory representations and strategies of the stakeholders to assert control and appropriation of the territory, and how these stakeholders defend their interests within this territory.

On the one hand, it is clear that cyberspace has become a relevant object of study for the field of geopolitics. Multiple actors have been empowered by the speed, ubiquity and anonymity provided by the low-cost and widely accessible technology for better—the multiple benefits to our societies— and for worse [Douzet 2014]. Cyberspace has indeed become a space of conflict and a strategic priority for many states. In 2016, at the Warsaw Summit, NATO allied nations recognized cyberspace as a new "operational domain in which NATO must defend itself as effectively as it does in the air, on land and at sea" [Brent 2019], thus following the United States and a wide range of countries that have identified cyberspace as a new military domain. The digital space is indeed used as a vector of attacks to perpetrate crime and conduct operations of sabotage, disruption, espionage or subversion [Rid 2013; Maurer 2018]. Cyberspace can therefore be regarded as a new dimension of most geopolitical conflicts and a new space of confrontation that lends itself to a geopolitical analysis.

On the other hand, understanding and representing the geography of cyberspace remains an ongoing challenge. If the notion of territory is at the heart of the scientific approach in geopolitics, cyberspace does not meet the usual criteria of the classical definition of a territory in geography [Lacoste 2003], in spite of the many metaphors describing it as a "global village", a "battlefield", or an "independent territory" [Douzet and Desforges 2018]. Nevertheless, just as with any other territory, cyberspace is a space organized by humans with its own geography, one that we need to comprehend in order to understand its dynamics, specific features – all to be able to analyse the strategies of the actors contributing in shaping this environment. This geography requires a dialogue between disciplines, associating geopolitics, computer science and mathematics to understand how to measure and represent different dimensions of cyberspace in order to identify its strategic facets.

We understand cyberspace to include both the Internet —i.e. the global interconnection of equipment for the automated processing of digital data — and the “space” it generates, a space of interactions and confrontations between multiple stakeholders. Cyberspace is sometimes represented as a structure composed of three –or more- superposed layers (the physical, logical, and informational layers) which are somewhat permeable and interact with each other. The physical layer composed of cables, servers and other physical equipment, is the material infrastructure of cyberspace, grounded in the physical territory. Therefore, it can be well understood and easily mapped with the traditional tools of political and physical geography. The two other layers are more challenging and require interdisciplinary work as geopolitical questions emerge from technical questions which in turn raise other strategic issues.

This paper focuses on the logical layer through an analysis of the structure of connectivity and the Border Gateway Protocol. The information layer has already been the object of much attention in the 2010s due to jihadist propaganda and manipulations of information in the context of elections, eventually leading to innovative cartographies of social networks and of the modes of content propagation [Howard et al. 2018 ; Limonier 2017]. The strategic dimension of the BGP architecture and data routing, however, has been given much less attention in the scientific literature.

The Border Gateway Protocol (BGP) which determines the routes the data takes, as we will see below, has been leveraged for many purposes:

1) by stakeholders to control the flow of information through routing traffic over specific paths
2) by countries to block the access to some contents or exclude some users from the Internet (going up to fully disrupting the internet) or for active malicious and strategic purposes such as hijacking traffic from other countries or attacking their infrastructure.

Because of its complexity and relative lack of transparency, the BGP architecture makes it possible for actors to develop indiscernible strategies and to largely mask their behaviours.

The article "Understanding the Network Level Behaviour of Spammers" [Anirudh and Feamster 2006] showed that traffic diversions have been used by spammers on the network. Since then, numerous studies [Vervier, Thonnard and Dacier 2015; Butler et al 2010] have revealed some inherent fragilities in BGP. Several countries have opted for a network architecture facilitating the definition of a BGP strategy. The goal of this study is to characterize objectively the different strategies adopted by the main decision-makers in a chosen region, to link them to current connectivity architectures and to understand their resilience in times of crisis. Our hypothesis is that there are clear bridges connecting the network architecture with the implementation of a BGP, that uncovers the strategy of stakeholders at a national level.

Several articles have explored nation state strategies involving BGP. Edmundson et al. [2018] elaborate a methodology based on traceroutes quantifying the importance of specific set of nations for the global routing. Karlin et al. [2009] describes on a coarse level the impact of the extra-territorial Internet routing and the associated risk in terms of sovereignty. Wahslich et al. [2012] develop a taxonomy of the AS-level connectivity which aims at understanding the role of the main actors of the connectivity in Germany.

We chose to focus on the case of Iran for two reasons. First, from a technical standpoint, Iran presents an interesting BGP architecture and holds a central position in the connectivity of the Middle East, the region of the world that has seen the largest growth in Internet penetration over the past decade [Gelvanovska, Rogy and Rossotto 2014]. Second, from a geopolitical point of view, Iran is a major actor in the Middle East and at the center of several ongoing geopolitical rifts. The withdrawal of United States troops from Iraq in 2011 allowed Iran to become a more important player in the region and the country clearly aspires to become an undisputed major regional power [Giblin 2018]. This desire involves consolidating domestically the stability of the regime while asserting its power on the regional scene.

Due to its geographical situation and its long history, Iran lies at a "crossroads of [different] worlds" [Wright 2000]. The country has a singular position in the Middle East. Geographically it is at the intersection of several geopolitical spaces: the Caucasus, Central Asia, the Indian peninsula and the Arabic sphere. Moreover, the Persian language and its culture are singular elements surrounded by the Arabic, Turkish and Hindu blocks. Iran is the main stronghold of Shiism, which is a minority in most of Middle Eastern countries beyond Iran, Iraq and Azerbaijan. Noticeably, this component isolates Iran. The succession of invasions and foreign interferences has led to the emergence of a strong desire for autonomy and a persistent fear for its national security. Iran is also an heir to a long imperial history that shapes its nationalism, a fiery pride that legitimizes its desire to be the dominant nation in the Middle East. The theocratic nature of the regime, along with the ongoing power struggle with the United States, Saudi Arabia and Israel and the precedent given by the StuxNet attack, have driven the Iranian government to consider the Internet as a strategic element that can be both a major source of risk but also a strategic asset to spread its influence. In this context, the organization of the networks on the Iranian territory can help to understand the different dimensions of the Iranian strategy.

Our observations make it possible to infer three ways in which Iran could have deliberately used BGP to achieve its strategic goals: the pursuit of a self-sustaining national Internet with controlled borders; the will to set up an Iranian Intranet to facilitate censorship; and the leverage of connectivity as a tool of regional influence.

## II) Materials and Methods:

### 1. Datasets

To support our observations, we used several datasets and statistical methods that we describe in this section.

1. We used graphs to represent the connectivity between the individual network operators - Autonomous System (AS) - connected to each other. We inferred the graph of ASes by using the path advertised by the routers running the Border Gateway Protocol (BGP) to update neighbouring routing tables. The largest source of publicly available BGP routing data are collected in 2019 by RouteViews[1] and RIPE Routing Information Service (RIS)[2] which aggregate BGP messages from BGP monitors at cooperating Autonomous Systems. We augmented the BGP announcements by adding relevant information like 1) the name associated with the AS, 2) the country where the AS is registered, 3) the number of IP address prefixes announced by the AS, 4) the number of times a connection has appeared on the routing table.

2. We used the Potaroo blog to get statistics about the number of prefixes and ASes associated with each country year after year[3]. GDP data and Internet accessibility statistics across the globe come from the World Bank website and are from 2017.[4]

3. We also used a dataset compiled in 2009 by the Berkman Center for Internet & Society to quantify the complexity of Internet connectivity (a notion that we define below) [Roberts et al 2011]. We have re-calculated this metric for the graphs we have derived.

4. We gathered the *AS relationships* [Dimitropoulos et al 2007] inferred by Caida Research center at UCSD which indicates the underlying economic forces that drive the evolution of the Internet topology and its hierarchy.

---

[1] http://www.routeviews.org/routeviews

[2] https://www.ripe.net/analyse/internet-measurements/routing-information-service-ris

[3] "BGP Routing Table Analysis Reports", Houston G. Blog, https://bgp.potaroo.net/

[4] World Bank Open Data. https://data.worldbank.org/.

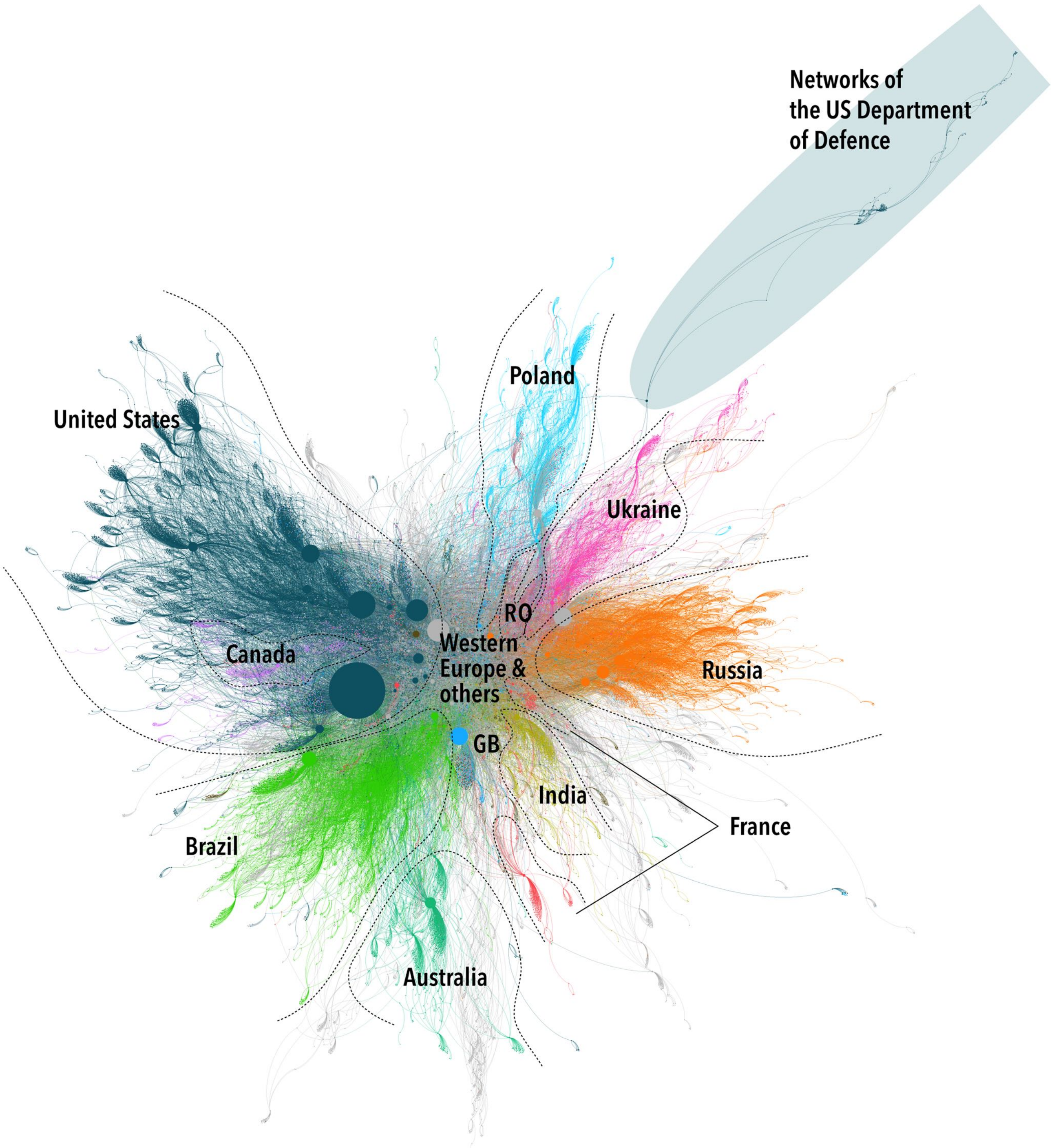

*Figure 1: Representation of all the ASes using the Force-Atlas 2 algorithm [Jacomy et al 2012]*

## 2. Analysis of the Internet connectivity architecture through BGP

The Internet is, as its name suggests, a network of networks. It is made up of around 70,000 ASes, each identified by a unique number (see Figure n°1). Each AS consists in a set of routers and communication links managed by a single administrative authority that decides which routing policies to apply. Each AS owns a set of prefixes. A prefix represents a set of contiguous IP addresses that are assigned by a regional registry representative of the ICANN to the AS[5]. The largest AS is AS335 and belongs to Level 3 Parent LLC; it announced 1,328,298,841 IPv4 addresses in May 2019. It is then followed by AS721, owned by the US Department of Defence, with 89,384,192 addresses. In contrast, some ASes contain a single computer only. Each autonomous system has full control and authority on routing inside its network and over the access policy applied to traffic transiting through it. However, an AS have to interact with its neighbours in order to exchange traffic with them. If we compare the Internet to the globe, autonomous systems, or ASes, can be considered as countries, each with its own legislation and separated by borders.

ASes use specialized machines called routers to exchange packets and inform their neighbors about their routing policies. Two routers belonging to different ASes but connected to each other communicate via the Border Gateway Protocol (BGP). Through BGP, each router announces to its neighbours in other ASes whether they can access an IP address while passing through the AS to which it is attached. The paths followed by packets are therefore defined by BGP, which can be described as the glue holding together the Internet [Robine and Salamatian 2014]. The structure of the links between ASes is defined logically by BGP and not uniquely according to the physical connectivity. For that reason, BGP defines the "topology" of the ASes through their connectivity structure, relationships, dynamics and defines by extension the shape of cyberspace.

To enable the access to IP prefixes that are not belonging to them, ASes have to purchase transit services. Although the arrangements between Internet Service Providers (ISPs) are complicated, there are commonly three types of possible relationships:

1. customer-to-provider (c2p) and provider-to-customer (p2c)

2. peer-to-peer (p2p)

3. sibling-to-sibling (s2s)

The first case corresponds to the scenario where one of the ASes is a customer who gains access to a transit AS (also known as Internet Service Provider, ISP) through a monetary contract. A p2p connection links two ISPs of equivalent size that agree to exchange freely some predefined amount of traffic, instead of defining two converse c2p contracts. Finally, a s2s link connects two ASes under the umbrella of a common administrative entity. A path connecting two prefixes is valid if all the ASes that the traffic passes through admit a connection of type 1, 2 or 3. When there are different possible paths, the AS determines which route to use to transport the traffic according to its own policy, which is unknown to an external observer [Luckier 2013]. The decision to announce a route and let traffic pass through an AS depends on the operator's commercial policy, its strategy, and the competitive environment he operates in as well as technical characteristics like delay or congestion. Thus, the path that packets take to move from one point to another may change according to the trade agreements and to the competition between economic or political actors. BGP is therefore a field of frictions between the different actors of the network.

[5] For Europe, Middle East and Central Asia the Regional Internet Registry (RIR) is RIPE that is based in Amsterdam, Netherland

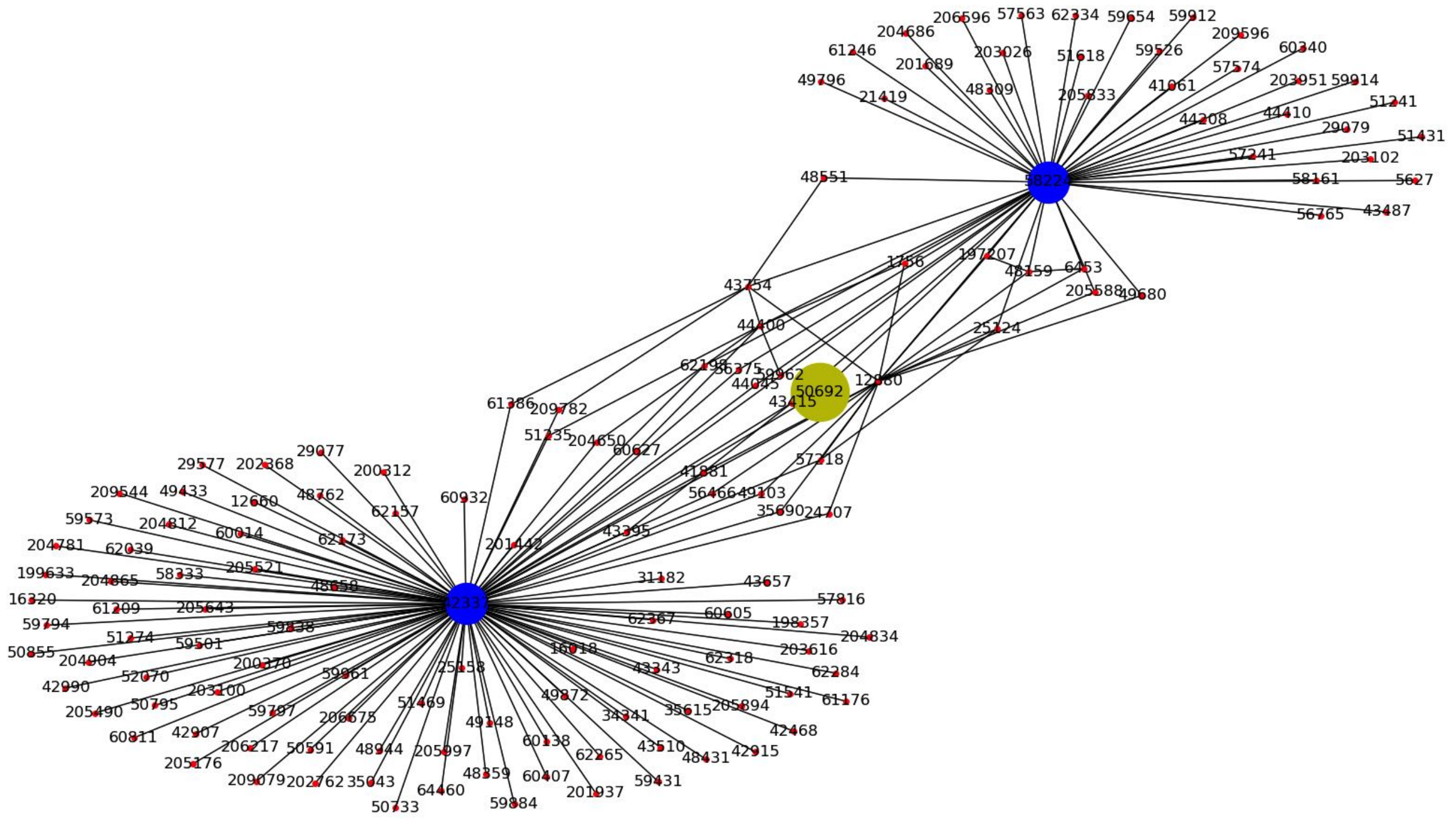

***Figure 2: Iran Kodro's AS (represented in yellow), its neighbours (in blue) and the neighbours of its neighbours***

The BGP actor policies results into BGP connectivity patterns and topologies that are represented by AS level graphs. The graph in Figure n°2 shows the connectivity patterns and topology for AS50692, an AS belonging to the Iranian car manufacturer Iran Khodro. Each circle indicates an AS that relays Iran Khodro's connectivity announcements. These graphs that are used by routing algorithms contain technical information that is difficult to interpret for non-specialists. We can observe that while each of these ASes relates administratively to a given state, many of them extend far beyond their national territories and may even be the aggregation of machines at different geolocations.

## III) Results : Iran, a case study

## 1. Iran's BGP Tree Architecture : a Strong Control at its Borders

We first looked at the architecture of Iran's ASes and the way they are connected to the rest of the world. With a population of more than 80 million and an Internet penetration rate above 50 per cent, Iranians make up a significant share of Internet users in the Middle East. As shown in Figure n°5, Iran consists in a total of 435 Ases in our inferred graphs. In terms of the number of ASes, Iran ranks 29th globally with 0.71 per cent (750 ASes) of the overall ASes allocated in the world. However, only 448 ASes, representing 68% of all ASes allocated to Iran, were advertised in the network and about 90 per cent of them contain less than 50 prefixes, which means that almost one third of all ASes allocated to Iran are not yet in use. More than 12,700,000 IP addresses are registered in *Iran*, ranking the country 32nd globally with about 0.35% of the whole IP addresses space in the world. Among these addresses, 98,5% IP are announced, which represents about 0.34 per cent of the entire announced IP addresses[6].

In Figure n°3, the AS architecture shows the major Iranian ASes (ASes that advertising at least 5 prefixes) along with their international neighbours and the importance of their connections. In this graph, we discard edges which are often not operated (because they are used as backup links or for private peering), *i.e.* edges which are announced less than 3 times in different routing tables. Domestic service providers are represented in green, American providers in blue, Europeans in red and other suppliers in the Middle East in yellow. The thickness of a line between two ASes indicates the betweenness centrality of the connection, i.e., the proportion of the shortest paths between all nodes of the graph that go through this link. The betweenness centrality measures the impact of disconnecting this link on the global connectivity of the network [Nan, Guan and Zhao 2008].

Figure n°3 provides interesting insights. Our first observation is the relative lack of direct connectivity between Iran and most of its neighbouring countries. For example, there is no direct connectivity from Iran to Saudi Arabia, Bahrain or Kuwait. While there is indeed a possibility of communication between these countries, it goes through intermediary network providers, mainly attached to the US or the UK. This *de facto* situation clearly results from a geopolitical situation which has led to minimal economic interactions and infrastructure development between Iran and most of its neighbouring countries (except Qatar, Oman and Turkey).

[6] https://bgp.potaroo.net

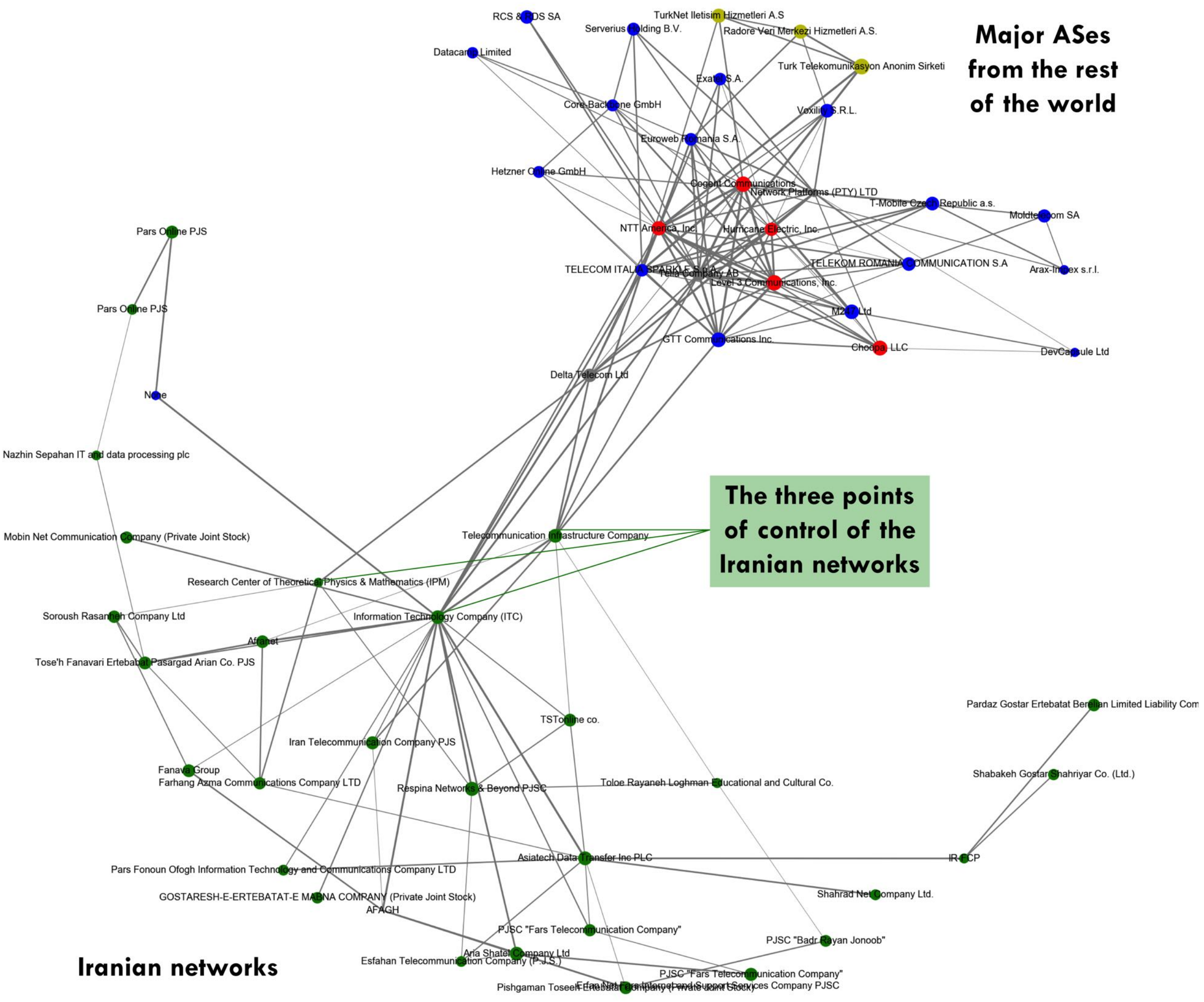

*Figure 3 : Simplified representation of the Iranian ASes*

Furthermore, there are only two Iranian ASes that connect most of Iran traffic to the rest of the world: the Information Technology Company (ITC - AS12880) and the Telecommunication Infrastructure Company (TIC - AS48159), which therefore play a key role at the border of the network. ITC is under the aegis of the Ministry of Information and Communication Technologies of Iran and the Telecommunication Infrastructure Company is directly affiliated to the ICT. This shows that Iran has built its architecture around the idea of a self-sustaining national Internet, connected to the global Internet by two highly controlled ASes.

The Iranian network appears to follow a tree. We can delineate (1) a trunk consisting of highly interconnected government owned ASes which lead the path to foreign networks, and (2) branches managed by private ISPs. While these branches are well connected to each other they do not have a great diversity of paths linking their traffic to the outside and they have to pass through the trunk. Such a backbone allows Iran to control the information exchanges with international network since all traffic goes through government ASes that may very well decide to stop it. There is a bottleneck between the Iranian Internet and the rest of the Internet; and the government ASes ITC and TIC play the role of gate-keepers that control the access to foreign content and decide what traffic passes through Iran.

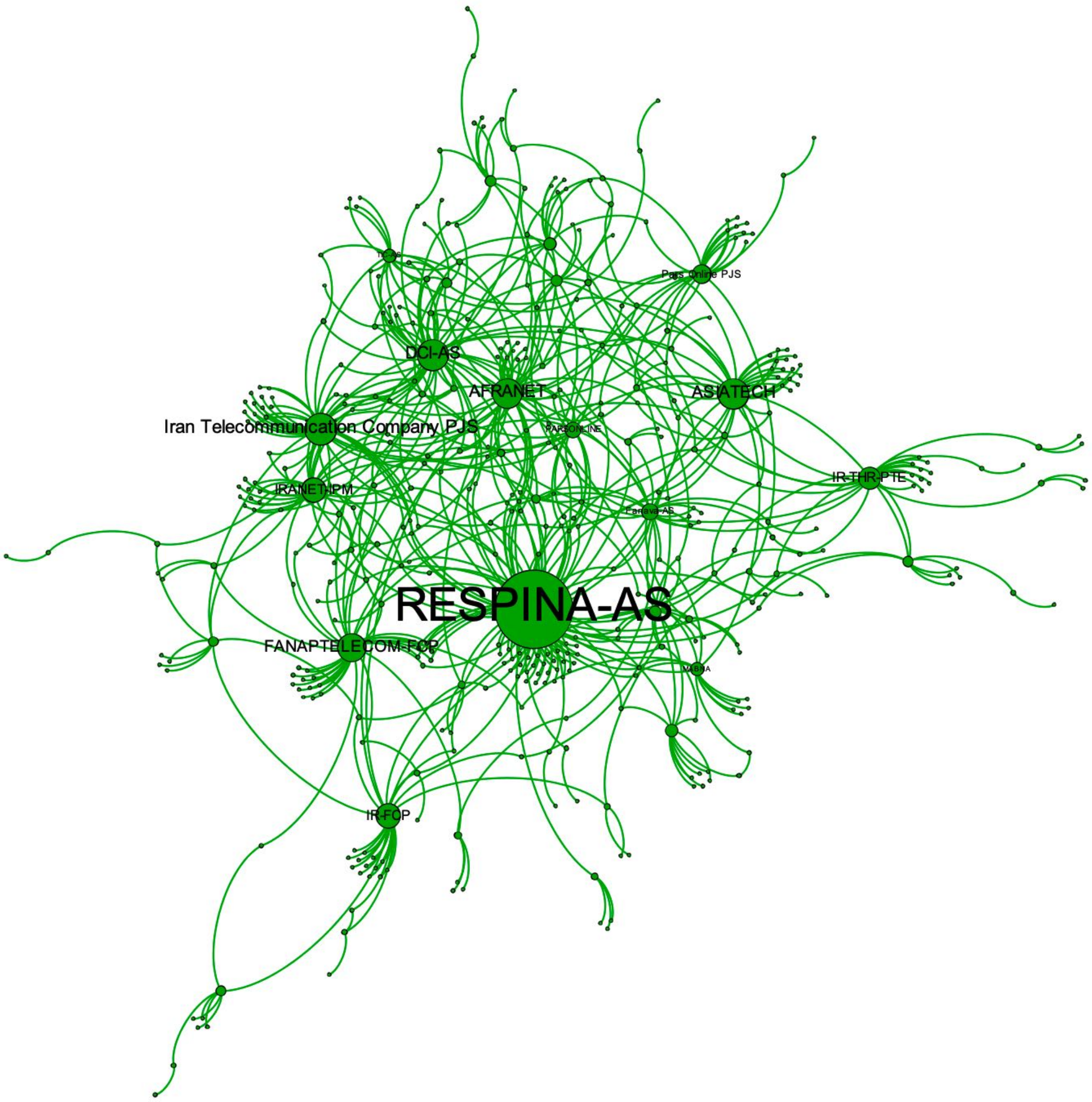

***Figure 4 : Zoom into the Iranian BGP landscape***

Zooming onto the Iranian AS ecosystem (Figure n°4), we can notice that the most central ASes are almost evenly disseminated within the network. The domestic network is therefore not centralized. Respina (AS42337) is the most central ISP, connecting to more than 100 ASes. We also notice that TIC is under-represented within Iran's connectivity graph despite the fact that it plays a fundamental role as a connection to the rest of the network as we have seen in Figure n°3, *i.e.,* internal ASes are not directly connected to TIC but rather connected through proxies that aggregate the traffic and eventually implement filtering (content censorship as will be described later).

As a first and simple comparison, we present in Table n°1 the number of external and internal BGP connections of Iran and other comparable countries in the Middle-East. We observe that Iran has a relatively larger proportion of internal connections, *i.e.* connections linking two ASes within the he same country, as opposed to Israel, for example, that has a larger proportion of external connections. This gives some leads into observing that the Iranian domestic network architecture is different from other countries and that this difference is likely caused by strategic decisions and design. While, the above descriptions give some elements for understanding the overall structure of the Iranian domestic network and comparing it to other countries', they are not sufficiently refined to provide a detailed comparative analysis with other countries. In the next section we will derive the complexity score to get a more precise picture.

| External connection | Internal connection | Total ASes observed | Country |
|---|---|---|---|
| 79 | 643 | 472 | Iran |
| 113 | 191 | 140 | Saudi Arabia |
| 113 | 180 | 109 | Iraq |
| 97 | 12 | 12 | Oman |
| 52 | 11 | 10 | Qatar |
| 40 | 136 | 132 | Lebanon |
| 128 | 18 | 22 | Bahrein |
| 24 | 41 | 37 | Jordan |
| 26 | 74 | 61 | Kuwait |
| 162 | 93 | 63 | Egypte |
| 15 | 51 | 49 | Afghanistan |
| 83 | 67 | 76 | U.A.E |
| 395 | 326 | 261 | Israel |
| 130 | 567 | 473 | Turkey |

***Table 1: External connections, or connections between two ASes are not located in the same country. Conversely, local connections are links within the same country***

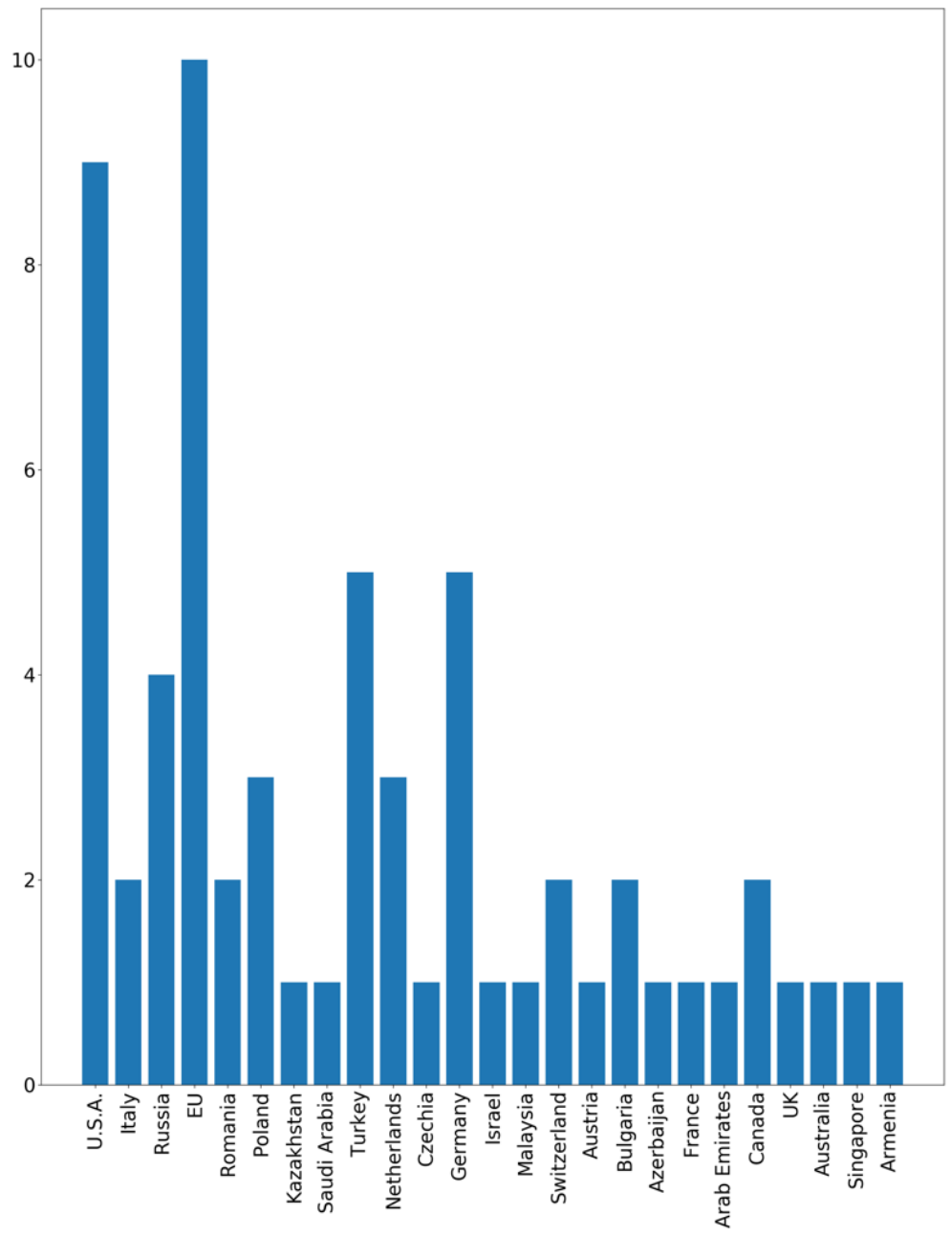

***Figure 5: Histogram of the nationalities associated with the direct neighbors of ASes registered in Iran at RIPE***

## 2. A Complex Network : low government control but strong resilience and opacity

In order to have a better understanding of the architecture and the dynamics of the network, we use the complexity score initially developed by Berkman Center of Internet and Society [Roberts et al 2011]. This metric captures the complexity of the network within a country by looking at the diversity of the announcements of IP addresses assigned to the country. A country where all or a major part of the available IP addresses visible from the outside[7] belong to a limited number of ASes will have a low complexity score. This means that internal actors have a limited choice of network providers. Inversly, a large diversity of IP addresses ownership by a large number of Ases is a sign of a more complex ecosystem within the country and the possibility of a larger set of potential routing paths through more providers to connect to each other or potentially to the global Internet if the AS connectivity structure allows.

As an illustrative example of a small network, let's suppose that a country wishing to implement a strong control of the Internet permits only one single AS to provide all IP addresses visible to the outside. This enables the country to exert a perfect control of the traffic, as it would be impossible to access the network without passing through this AS. but it would entail a very fragile network both internally and externally as this AS would become a single point of failure.

[7] IP addresses visible from outside might be accessed from outside the domestic network, through announced BGP paths.

1. A more complex network will be more resilient as the diversity of existing paths and visible address makes it possible to circumvent the potential points of failure. T
2. To achieve a more complex and therefore more internally resilient domestic network requires to increase the number of domestic ASes and enable their IP addresses to get access to the global Internet through a richer set of alternative paths. However, this comes in contradiction with the wish to control all traffic, since controlling a larger number of AS and outgoing paths is more difficult. The complexity score developed by the Berkman Center comes as a way of quantifying this trade-off between internal resilience and control.

The Berkman Center proposed another intuitive metric: the control value [Roberts et al 2011]. This metric leverages the notion of "points of control" defined as the minimal set of ASes needed to connect 90% of advertised IPs in the country to the external world. The proportion of ASes of the country in the points of control set define the control value. For example, if a country A possesses 300 ASes and if it takes 30 ASes to connect 90% of the IP addresses announced, then the size of the points of control set is 30 and the control value is 30/300=10%. The lower the control value, the greater the centralization of the network.

These two scores: complexity and control value, are somewhat redundant. The control value focuses on the accessibility to the Internet while the complexity score focuses more on the internal complexity. Looking at these two metrics gives us methodological tools to compare different countries' network architecture by evaluating where they stand on the control vs. resilience trade-off. The control value is a measure of the concentration of IP addresses within a small number of ASes and by extension of how concentrated the architecture of the routing is. The complexity, on the other hand, measures the routes data actually travels. Complexity is about control and quantifies how the country's users may have their traffic exposed to observation, manipulation and disruption. Both values are complementary as is shown on Table n°2.

We present the complexity metric calculated by the Berkman Center for Middle Eastern Countries in 2011 along with our own calculations for 2019. Our sources of data are slightly different from the one used by the original paper: we have used the full BGP connectivity graphs while [Roberts et al 2011] used CAIDA ISP relationship data alone. Table n°2 shows that, compared to other countries in the Middle East, Iran has a highly complex network with a great number of interactions between its ASes.

| | Complexity | | Control value | | Number of ASes | |
|---|---|---|---|---|---|---|
| Country | 2011 | 2019 | 2011 | 2018 | 2011 | 2019 |
| Iran | **3.82** | **3.75** | **2%** | **34%** | **96** | **437** |
| Saudi Arabia | 3.74 | 0.43 | 5% | 10% | 66 | 139 |
| Iraq | 6.46 | 4.93 | 75% | 55% | 4 | 107 |
| Oman | 1.06 | 0.05 | 50% | 25% | 2 | 12 |
| Syria | 0.85 | 0.00 | 33% | 50% | 3 | 2 |
| Bahrain | 10.20 | 0.26 | 22% | 37% | 18 | 19 |
| Kuwait | 4.70 | 0.52 | 20% | 17% | 30 | 61 |
| Egypt | 1.25 | 0.04 | 8% | 9% | 36 | 58 |
| Afghanistan | NA | 4.14 | NA | 52% | NA | 46 |
| UAE | 0.58 | 0.31 | 20% | 20% | 8 | 65 |
| Turkey | 2.72 | 2.67 | 1% | 7% | 226 | 450 |
| Israel | 3.24 | 2.41 | 2% | 10% | 165 | 251 |
| Qatar | 1.55 | 0.02 | 40% | 29% | 5 | 9 |
| Lebanon | 11.99 | 7.81 | 22% | 42% | 32 | 133 |

***Table 2 : Complexity of the Middle Eastern Countries on 2011 and 2019***

More interestingly, Iran —like Turkey— has maintained a high level of complexity from 2011 to 2019, while other Middle-Eastern Countries have drastically reduced the complexity of their networks over the past 8 years, showing that the Iranian strategy of development of its internal network is significantly different from the other countries'. This tendency is exacerbated in the Gulf countries where we see that most values are hovering around the value of 0.5.

The control value scores (see Table n°2 and Figure n°6) confirm the above results. In 9 countries out of 15, the control value is below 30%, meaning means that 90% of the traffic circulates through less than 1/3 of the network. In Saudi Arabia, Turkey, Egypt and Israel, the level of centralization is particularly high, with a control value close to 10%. This architecture concentrates traffic onto a few ASes, making the network vulnerable to traffic congestion, potential failures or cyberattacks toward these few ASes. This architecture however does facilitate control of content and users within the boundaries of the network. Iran, on the other hand, has a much higher complexity score and therefore a much lower level of centralization, with a much higher control value close to 35%. This means that the connectivity inside Iran is better distributed among the existing ASes. However, it does imply that there are more Ases connected to the global Internet.

For the case of Iraq and Afghanistan, we observe control values over 50%, which reflects the lack of strong central authority that can control and shape the architecture of the domestic network. There are therefore mutliple points of connection to the global Internet. It is noteworthy that even if Israel and Turkey have relatively large complexity score, their control value is much lower than in Iran, showing that these two countries have a stronger control grip over their national network infrastructure, and this despite the fact they have a larger number of connections to foreign countries.

These trends reflect different representations of Internet Sovereignty and different strategies of control. In Iran, the emphasis is put on the control at the border of the domestic network, while there are more alternative paths for traffic flow within its borders. Whereas most Middle Eastern countries have obviously increased the centralization of their domestic networks to better control Internet users. Most domestic traffic is forced through a few controlled ASes, with the risk of creating domestic congestion and low resilience. Iran instead has managed to ensure a robust internal connectivity while maintaining the ability to isolate its network from the global Internet, in a way that does not imply congestion into central ASes.

The above discussion shows that Iran has succeeded in building a national network that reconciles two seemingly incompatible properties. On one hand, Iran domestic network is highly resilient with a relatively large number of ASes and a rich ecosystem of internal paths. On another hand this network is strongly under control with all the outgoing traffic flowing through two main government owned ASes and a relatively low control value. Iran has therefore managed to isolate its network from the global Internet while ensuring a robust internal connectivity, in a way that does not imply congestion into central ASes. The bottleneck therefore occurs externally.

This implies that the Iranian network could completely cut off access to the rest of the Internet for Iranian users and isolate its network without modifying its internal state. Moreover, Iran, through its rich domestic network, can modulate the level of disconnection according to its interests and strategic objectives. A complete disconnection of the Internet comes with high collateral economics costs but partial or temporary disconnection could bear some advantages. From this perspective the Iranian network architecture is becoming similar to the Chinese one, with the notable fact that this evolution has happened relatively quickly.

In contrast, other Middle Eastern countries have strongly decreased the complexity of their domestic network in order to achieve a better control of it but at the cost of a higher risk of disruption and lower resilience that can results from domestic congestion, accidental failures or, worse, cyberattacks.

An analysis conducted by Mahsa Alimardani demonstrated the clear impact of the Stuxnet attack — believed to be a cyber weapon jointly developed by the United States and Israel against the nuclear facilities of Natanz —on the sensitivity of Iran to the resilience of its network and by extension the creation of new ASes [Alimardani 2019]. The ongoing international embargoes in the past decade have also driven the Iranian government to develop the Iranian National Information Network project[8] that can work as an Intranet separated from the global Internet. Moreover, the clearly declared willingness of the Iranian government to protect its national Internet from foreign interference and the censorship implemented in Iran are all strong signs of the existence of a strategy of control that might motivate the previously described architecture. Over the past ten years, Iran has indeed increased its AS fleet at great speed and far ahead of its competitors in the Gulf. Figure n°7 shows the evolution of the number ASes in Iran from 2009 to 2018 (in pink) and the share of ASes announced in Iran compared with the rest of the World (in green). Managing an AS being a relatively technical job that requires a good level of expertise. This evolution therefore demonstrates that Iran has been able to gather the economic and technical resources necessary to manage this high number of ASes.

In this section we have demonstrated how Iran's BGP architecture can facilitate control at the border. The following section will document how BGP has obviously been actively used as a tool of censorship by the regime.

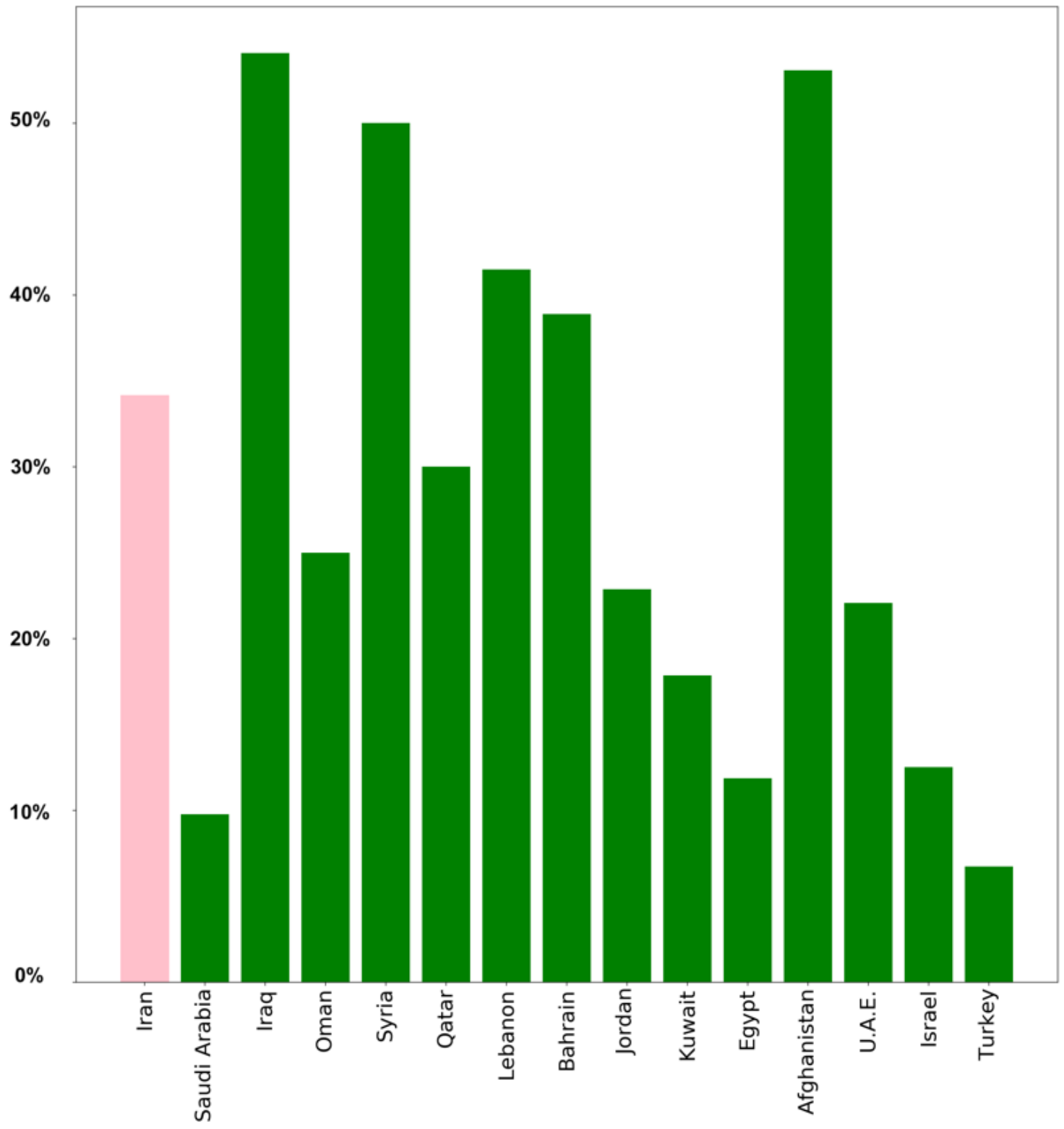

***Figure 6: Control value of the Networks in Middle Eastern Countries in May 2019***

[8] http://www.rrk.ir/Laws/ShowLaw.aspx?Code=1640. Accessed October 15, 2019.

### 3. BGP as a Tool of Censorship: The Dream of a "Halal" Internet

As the Iranian authorities work towards the creation of a "halal" Internet —meaning a domestic Intranet— it implies the establishment of a particularly sophisticated censorship system in Iran. As of 2012, an average of 27% of internet sites were blocked [Shaheed 2014]. If we look at the details of the blocked domains, we note that a large majority are news media and websites dealing with human rights issues. The most censored category of websites is unsurprisingly pornographic content. It is interesting to note that more than 50 per cent of the most visited websites across the world are not accessible to Iranian Internet users [Tor Blog 2011].

It is also very important to consider that content access control and censorship in Iran is not only caused by the actions of the Iranian government but increasingly it is implemented by foreign companies fearing to be in infraction with the Office of Foreign Assets Control (OFAC) of the American administration and implementing, sometimes even when waivers exists, nation scale bans to access their contents or services. Some examples of these content access control that have been imposed, are IEEE [Brumfiel 2004] in 2003, that was reversed in 2004 and is being reimplemented in 2019, Amazon Web Service in 2019[9], GitHub [Raadi 2019] in 2019, etc. While censorship have been implemented through traditional web filtering techniques, like DNS poisoning [Levis 2012], i.e., giving wrong answer to some DNS requests making some websites inaccessible, or web proxy filtering [Chen et al. 2010], more advanced techniques like Deep Packet Inspection (DPI) [Bendrath et al. 2011] have been deployed. However, the existence of embargoes toward Iran have reduced access to these technologies running on high speed links. And while Chinese companies are very active in the Iranian telecommunication market —in particular Huawei and ZTE[10]— , these companies are not selling filtering technologies even if China is a technological leader in this domain. The ever-increasing links speed has made difficult for Iranian censorship to follow up. Meaning that traditional censorship had been distributed into the network close to customers and the responsibility to implement it is given to Internet Service Provider. Nonetheless, this approach does enable fast controlling action as adding a new filtering needs the cooperation of all ISPs and take time to propagate. But, Iran had, as an essential strategic goal, the possibility of global communication "kill switches" that would enable from a central point to disconnect all Internet connections or even mobile communications. Meaning that they would like to have solutions of immediate censorship of contents. The specific domestic network architecture we describe above, along with mindful BGP manipulation provide the tools for achieving this goal.

[9] Center for Human Rights in Iran, "More Iranians Forced to Rely on Unsafe Online Hosting after Amazon Ban". https://iranhumanrights.org/2019/08/more-iranians-forced-to-rely-on-unsafe-online-hosting-after-amazon/. Accessed October 15, 2019.

[10] Su, J. "Analyst: China's ZTE Shuts Down After U.S. Tech Ban Over Iran Sales", *Forbes*, 9/05/2019. https://www.forbes.com/sites/jeanbaptiste/2018/05/09/analyst-chinas-zte-shuts-down-after-u-s-tech-ban-over-iran-sales/#421218d64b0a. Accessed October 15, 2019.

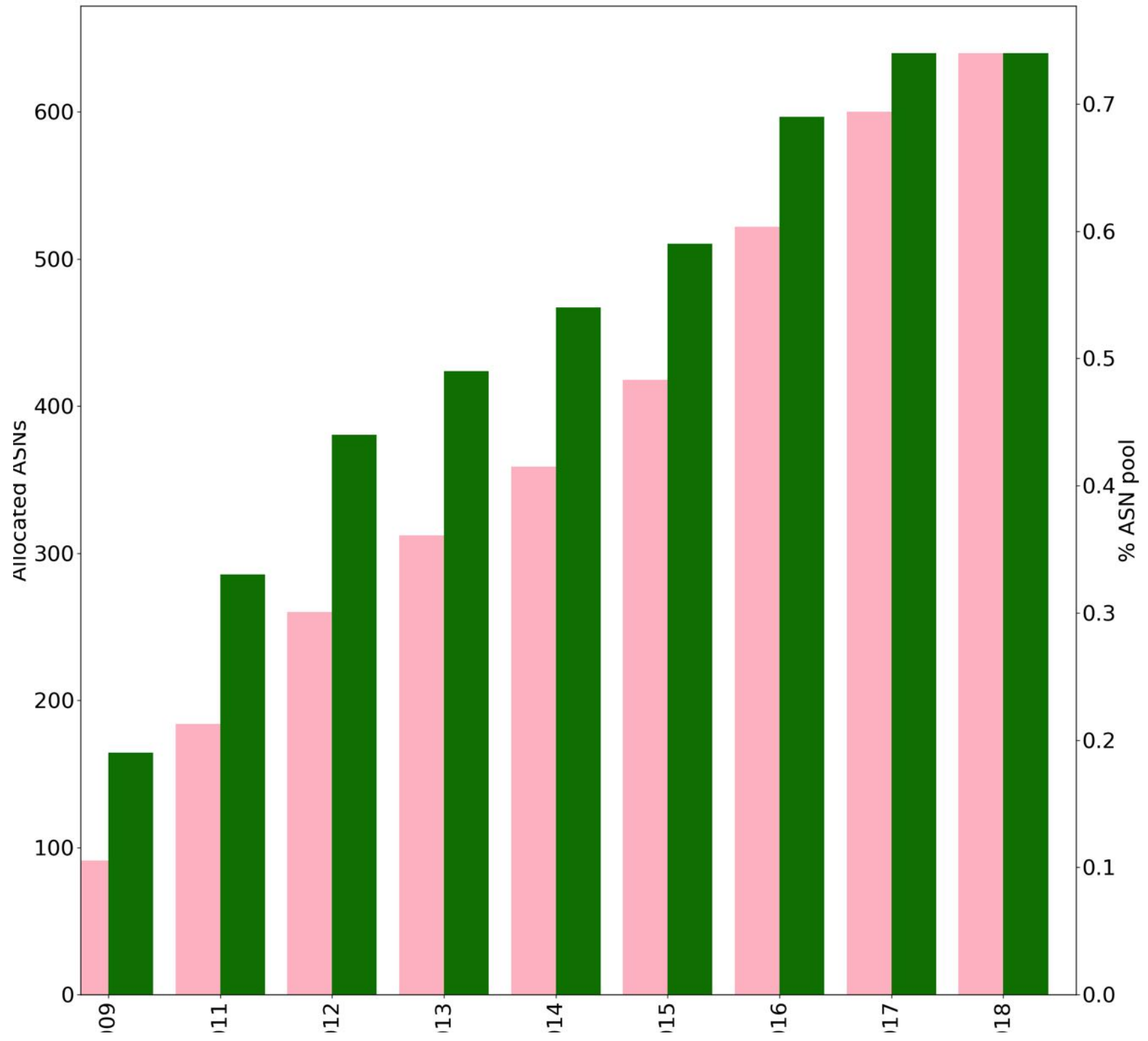

***Figure 7: Evolution of the number of ASes registered at RIPE in Iran and share of Iranian ASes among all countries.***[11]

In the forthcoming we will describe how the BGP architecture has been concretely used for censorship and also as an active tool for interfering with traffic. There are globally two classes of AS-related incidents : outages and hijacks. An outage happens when a prefix is no longer announced by any AS. It usually corresponds to a technical problem which is symptomatic of the fragility of an Internet Service Provider but the past few years have seen a rise in country-wide Internet Outages caused by national censorship [Dainotti et al 2011].

A hijack is the illegitimate takeover of prefixes by corrupting Internet routing tables across the graph. The traffic then follows paths that it should not be taking and transits through a new AS. This allows the new AS it crosses to analyse the nature of the traffic or to delete it.  In practice, hijacks are not all malicious. They might result from unintentional misconfigurations, inducing a change on the path data follows over Internet. Hijacks are frequently unwanted and resulting from misconfiguration, but they are also used with censorship aims.

---

[11] *Ibid.*

We show in Figure n°8 the relation between the number of ASes in different countries and the number of BGP related events, outages and hijacks, that has happened in this country. As can be seen there is roughly a direct relation between the number of ASes in a country and the number of observed outages. This is resulting from the fact that most of BGP event are configurations errors and the rate of these errors happening is roughly constant over all ASes regardless of their country. However, looking at the position of Iran in this Figure, we can observe that Iran is above the 95% regression confidence line, *i.e.,* the likelihood that Iranian BGP events can be explained similarly to the bulk of other countries as random configuration errors is less than 5%. Only two other countries are similar to Iran in this respect, Indonesia and Colombia. But both these countries have a large number of maritime cables (coming from the Caribbean Sea for Colombia), and have mainly witnessed outages resulting from issues in these cables (accidental cable cuts, *etc.).* Iran is therefore over-represented in BGP events with respect to the size of its AS eco-system and this is a sign of Iran « mis »-using BGP for specific purposes.

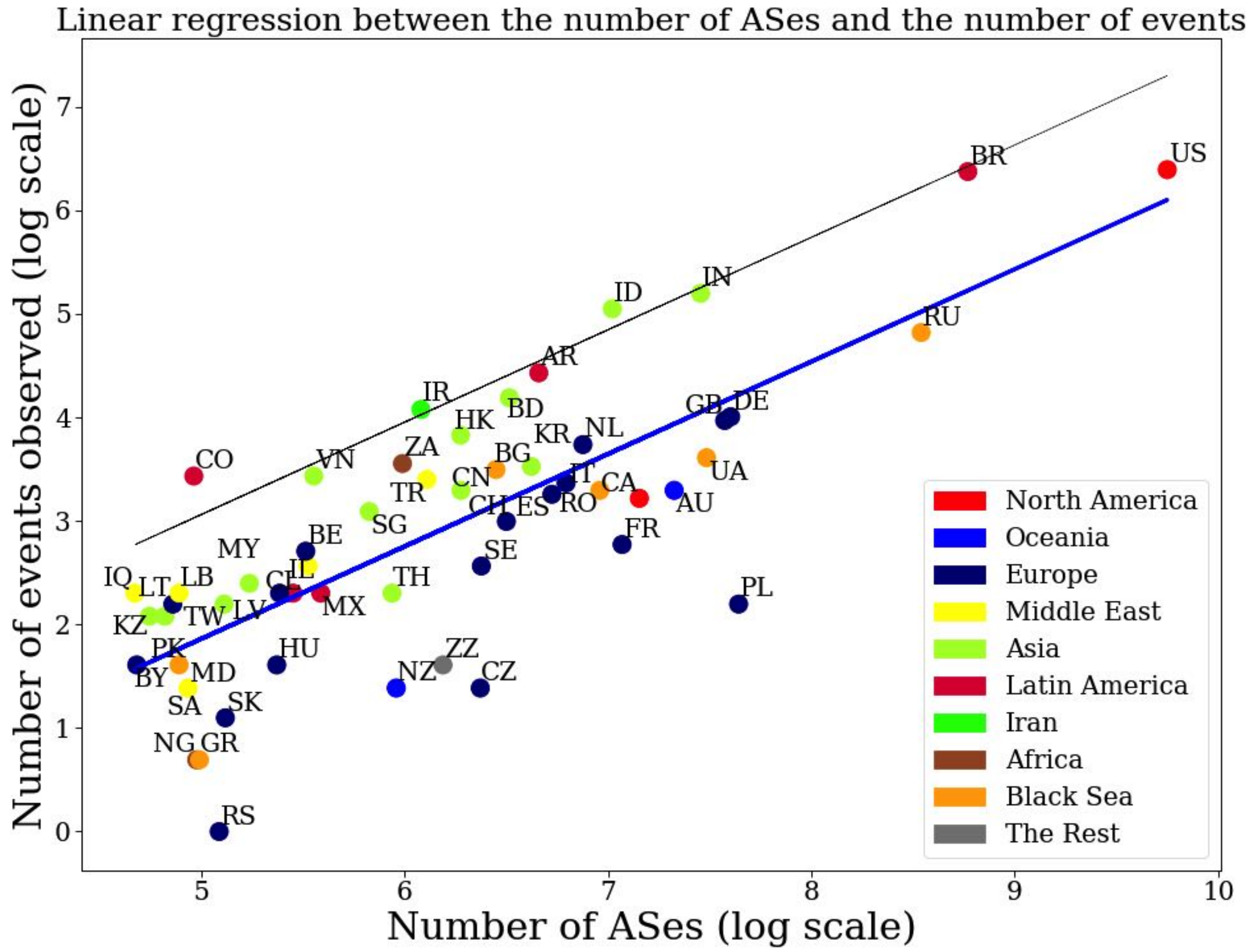

***Figure 8: Linear regression between the number of ASes and the number of BGP events. The different colors represent different regions of the World.***

A focus on Iranian outages is particularly instructive of the Iranian strategy [Abdelberi et al 2014]. in January 2017, Iran had already adopted a holistic approach to censorship through the systematic elimination of any outgoing connection via BGP tampering and a suspension of the ASes' outgoing traffic[12]. This incident is not isolated. Iran experienced a wave of protests against the regime between December 2017 and January 2018. On January 1, 2018, BGPstream[13] detected that more than 44 per cent of the Iranian prefixes were no longer accessible and that a majority of Iranian ASes had disappeared from the BGP graphs, all of this while demonstrations were in full swing in the streets of the largest cities. By disabling a population's access to the Internet, Iran showed its aptitude in conducting a sophisticated form of Internet censorship. While the straightforward approach usually consists in physically disconnecting critical infrastructures, Iran's control over its network allowed a more elaborate approach based on BGP to disrupt the routing of packets. On June 26, 2019, more than 80% of Iran's ISPs were disrupted and disconnected. A more thorough BGP-level analysis showed us that around 8pm, ITC AS (AS 48159) stopped announcing most of its prefixes externally (which corresponds to 26% of all the Iranian prefixes)[14]. This led to a complete reshaping of the Iranian network with specific ASes reacting by changing their associated path to certain prefixes. Most users within the country started noticing slow speeds and disruptions to the overall Internet. Even more interestingly, some users on Twitter described the network as a "True National Internet"[15] as connections to local services were not as severely affected as connections to the global Internet[16].

A very well documented case of this is BGP hijack by Pakistan Telecom[17] of You Tube in 2008. In January 2017, a similar event happened in Iran when the Iranian ITC took part in a hijack of prefixes that contained pornographic websites[18]. While the BGP hijack announcements were meant for the Iranian network solely, they got out of Iran because an ITC configuration error. The accidental announcement outside of the Iranian network spread the hijack throughout the network and suspended the activity of the pornographic websites on the entire Internet. This generated a reaction from the BGP traffic

---

[12] Reported by BGP Stream website as Event #2110094. Accessed July 26, 2019. https://bgpstream.com/event/210094; see also https://netblocks.org/reports/widespread-internet-disruption-in-iran-amid-geopolitical-crisis-3AnwGkB2 Accessed August 10, 2019.

[13] https://bgpstream.com/ Accessed August 10, 2019.

[14] https://ioda.caida.org/ioda/dashboard#view=inspect&entity=asn/48159&lastView=overview Accessed August 10, 2019.

[15] https://twitter.com/Pouyan_01001010/status/1143989760299585541 <Accessed August 10, 2019>

[16] https://bgpmon.net, https://ioda.caida.org/ioda/dashboard#view=inspect&entity=asn/48159&lastView=overview Accessed August 10, 2019.

[17] https://www.ripe.net/publications/news/industry-developments/youtube-hijacking-a-ripe-ncc-ris-case-study

[18] https://dyn.com/blog/iran-leaks-censorship-via-bgp-hijacks/

monitoring the services, such as BGPmon[19] and Dyn[20], which led to a correction of these advertisements. In July 2018, Iran hijacked the Telegram application traffic from all over the world and let it pass through Iran[21]

These examples illustrate how Iran « mis »-used BGP as a tool to intervene on the Internet and to control its character. In recent years, Iran has developed a highly opaque Internet which facilitates outages on a national scale such as the one we have just described. All in all, the control of the exit points of network makes it possible to hide most hijack operations within the network [Madory 2017]. For Iran, the controls at the borders allow not only to regulate the content entering the country but to manipulate the transit of the requests as well. The high complexity induces an extra-layer of *thickness*, this means that the traffic within the border appears to be *foggy* to an external observer. In the absence of monitors located within the country's network, getting a precise idea of the dynamic behind the route evolutions gets very tricky. Most of the major censorship events originating from Iran such as the ones described earlier were visible only due to a configuration error that spread to foreign ASes.

## 4. Connectivity as a Tool of Influence ?

During its long history as an empire, Iran has leveraged its geopolitical position at the intersection of several major geopolitical ensembles. This position along with its cultural aura has given Iran an influence that goes beyond its current borders and immediate neighbors. The large-scale conversion of Iran in 16th century to Duodecimal Shiism, a minority branch of Islam, has also been a major tool of influence. However, during the 19th century Iran lost most of its territory and influence in Caucasus (through the Golestan Treaty) and in central Asia (through Turkamenchay Treaty) [Rashidvash 2012]. During the 20th century the independence of Bahrain strongly undercut Iran's influence in the Persian Gulf. The 1979 Revolution resulted into a major change of the Iranian strategic environment. All Western allies of the Shah disassociated themselves from the Islamic Republic of Iran. The Iran-Iraq war made the country even more isolated with only Syria openly backing Iran. The decade following the end of the 8 year long Iran-Iraq war in 1988 —with the invasion of Koweit by Iraq followed by the first and second Gulf war, the Afghanistan war in the aftermath of the 9/11 attacks and the end of the Soviet union— put Iran in a favorable geopolitical situation with all the major risks in its direct neighborhood mitigated by US led coalitions.

---

[19] https://bgpmon.net

[20] https://dyn.com/monitoring-analytics/

[21] https://www.cyberscoop.com/telegram-iran-bgp-hijacking/ Accessed August 10,2019.

During this period Iran began to improve its domestic infrastructures and in particular its telecommunication backbone [Barzegar 2008]. The improvement came with the deployment of a large-scale fiber optic network, first between different large cities and thereafter even in rural regions. Based on this network, Iran developed a fiber optic industry and an expertise in deploying large distance cables[22]. This investment yield into the availability of a dense fiber connectivity networks from the west to the east and from the south to the north of Iran, putting the country in a strategic position to provide network connectivity to its neighboring countries and beyond.

Currently two major international cables cross the Iranian network: Europe-Persia Express Gateway (EPEG) and Trans-Asia-Europe (TAE). In addition, Iran has also access to the Fiber-Optic Link Around the Globe (FLAG) cable on its south via a direct connection to the UAE. The TAE cable runs from Azerbaijan Turkmenistan through the northern part of Iran, with an extension to Georgia and Azerbaijan. Yet it has been mostly used to provide connectivity between Turkmenistan and Azerbaijan

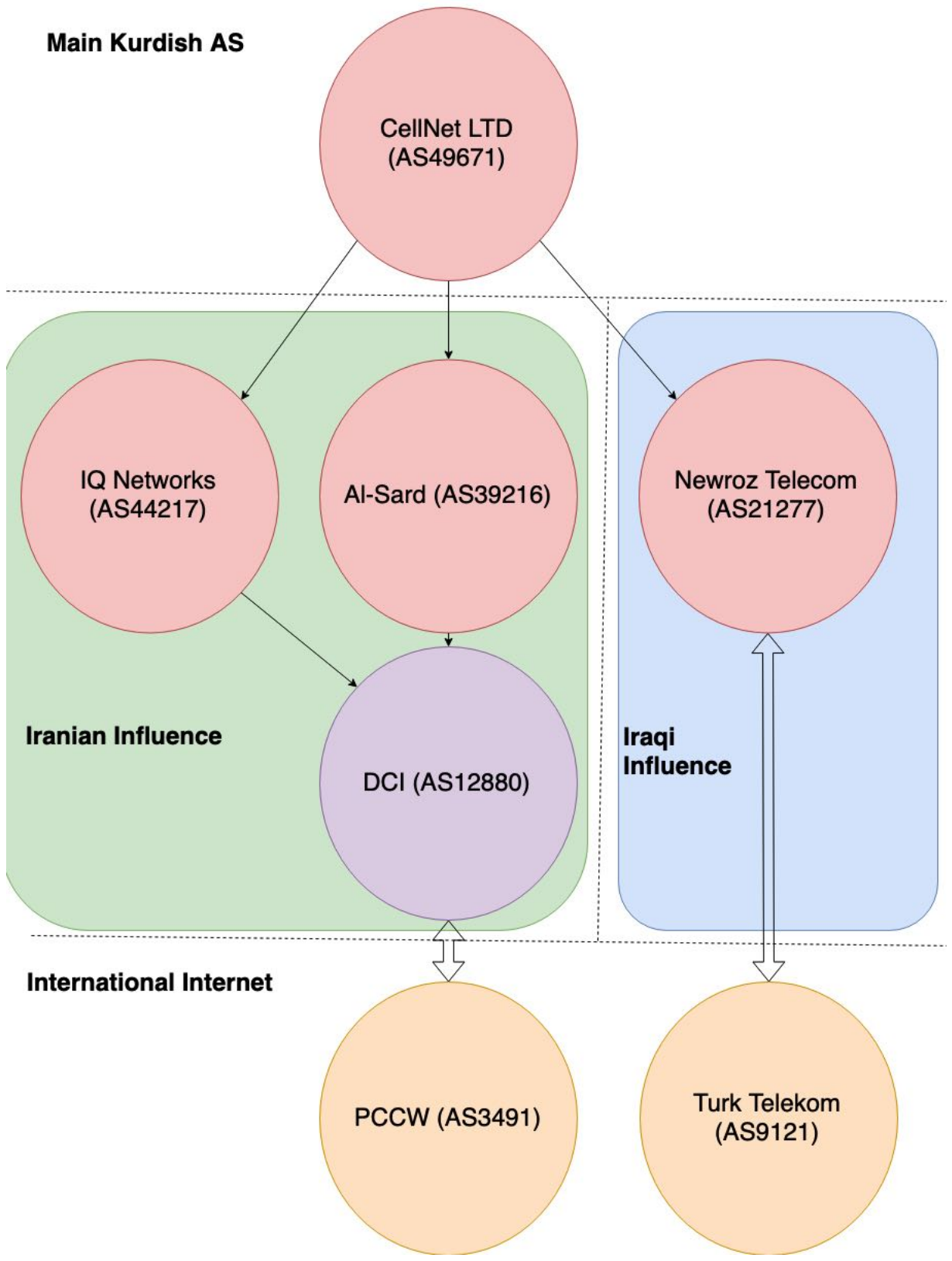

***Figure 9: Representation of connectivity among the Iraqi Kurdish ASes Al-Sard and IQ Networks, both using the Iranian AS ITC as an intermediary towards the rest of the Internet [Cowie 2010]***

at a maximal capacity of 2.08 Gbps. The EPEG cable with a total length of approximately 10,000 km,

---

[22] http://www.fiberopticom.com/news/iran-will-deploy-a-14-000km-fibre-optic-networ-21105894.html Accessed August 14, 2019

goes from Frankfurt —across Eastern Europe, Russia, Azerbaijan, Iran and the Hormuz straight— to Barka, Sultanate of Oman, where it is connected to a rich network of maritime cables. This cable is owned by a consortium of four carriers: Cable & Wireless, Rostelecom, Omantel, and TIC (Telecommunication Infrastructure Company of Islamic Republic of Iran). The maximal capacity of this cable is 3.2 Tbps but its current operational capacity is 500 Gbps[23]. This cable is very interesting from a geopolitical perspective as it is currently the only viable Internet traffic transit route path from East Asia to Europe that represents an alternative to crossing the Red Sea and Suez Canal. It is also shorter in length and can decrease by 10 msec the delay between Tokyo and Frankfurt. It is noteworthy that the cable became operational in 2013, only after two years after the Memorandum of Understanding[24] establishing it was signed, showing the maturity of the Iranian fiber network. This cable is currently operational, and its bandwidth gradually increases. There are ongoing discussions with Qatar to provide the country with an Internet connectivity that will not depend on other Gulf countries that are currently embargoing Qatar. Figure n°3 shows the central presence of Delta Telecom Ltd (AS29049), the Azeri entry point to EPEG and TAE, and therefore the importance of the EPEG and TAE cables.

The deployment of International cables in Iran has multiple benefits. First, it provides Iran with the international connectivity it needs. Moreover, this investment brings direct economic benefits to Iran Telecommunication Infrastructure Company (TIC) but also indirect economic growth to the country through the development of a digital ecosystem (*e.g*, datacenters, access infrastructures crossing Iranian Ases, engineering services, *etc.)*. The third benefit is strategic. The traffic flowing through EPEG is, for a notable part of the path, under the direct control of Iran, making it possible for the country to observe, monitor and interact with the data. Even if data encryption may prevent access the full content of the traffic crossing Iranian territory, metadata remain accessible. Moreover, the ability to interact with the traffic (*i.e.* block or disturb) gives an edge to a transit actor. This strategic advantage creates an incentive for countries to deploy Internet terrestrial cables through their territory and attract international traffic.

---

[23] https://www.vodafone.com/business/carrier-services/connectivity/submarine-terrestrial-cable/EPEG Accessed August 14, 2019.

[24] https://www.tic.ir/en/news/1600/Memorandum-of-Understanding-signed-at-the-sideline-of-EPEG-quadrilateral-agreement Accessed August 14, 2019

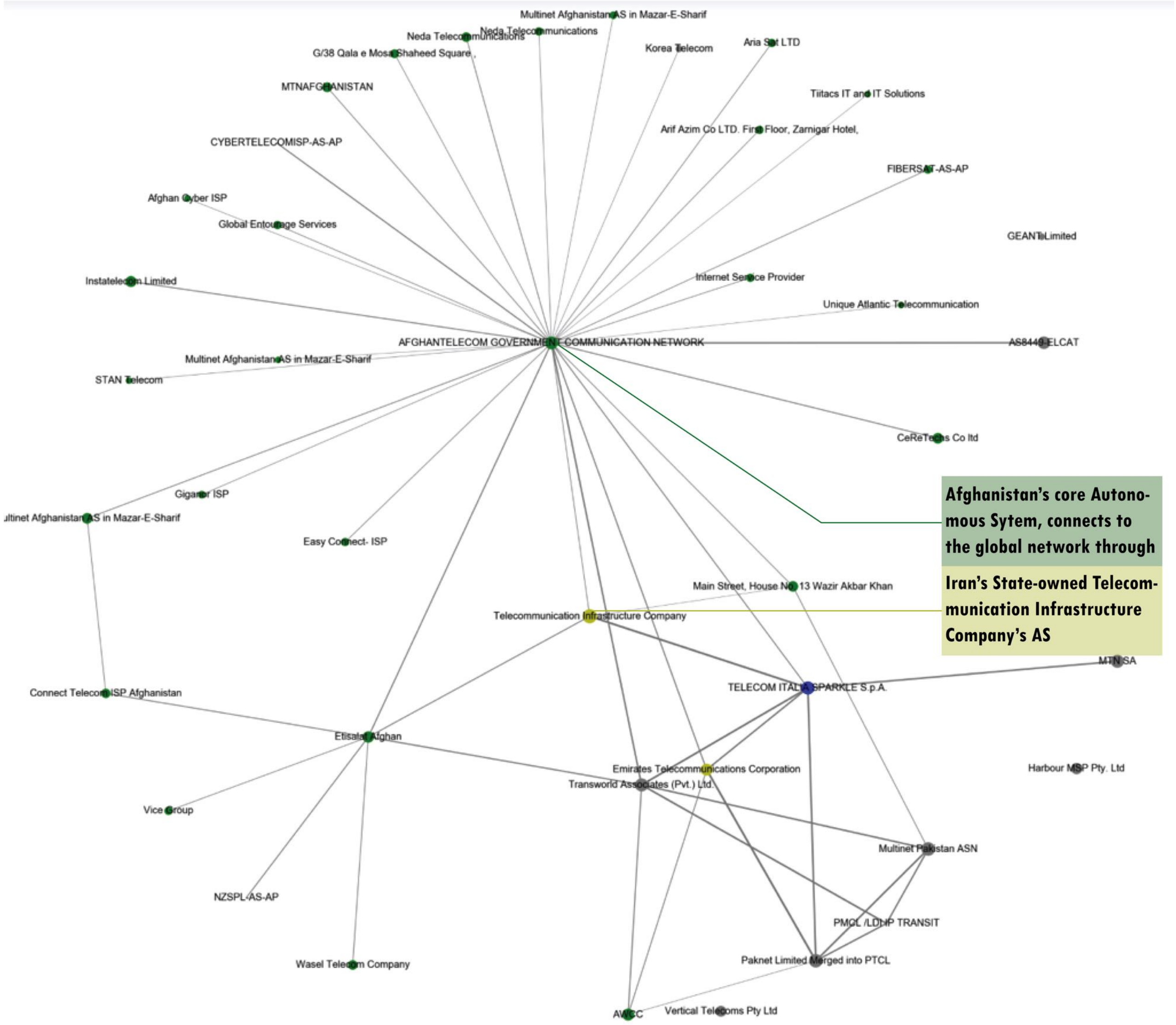

***Figure 10: Reduced representation of the Afghanistan graph in February 2018***

Iran has leveraged on its domestic network and international connectivity to attract traffic from its neighboring countries. The Iraqi Kurdistan has depended on the connectivity of Iran for its access to the Internet between 2010 and 2012. This region in the north of Iraq gained a large level of autonomy from Baghdad after the Second Gulf War [Gunter 1993] and decided that its Internet traffic should through Iran. This decision was partly motivated by the desire to avoid transiting directly through Turkey and to follow a geographically closer path. This resulted into the emergence of new paths originating from Kurdish ASes, going through Iranian Ases to access the global Internet. We show in Figure n°9 some of these paths. The situation has since evolved and the majority of the Kurdish traffic transits now through Nowruz, an Iraqi AS, and Turkey. Yet part of the traffic still transits through Iran to access Azeri ASes[25]. The evolution of routing in the Iraqi Kurdistan sheds light on how the geopolitical context shapes connectivity and reciprocally how the AS-level connectivity provides can be used as a tool in power relationships within the region.

Until the beginning of 2018 Afghanistan, another neighbor of Iran, was in a situation of dependency on Iran connectivity. Its network highly centralized around the AS Afghan Telecom AS59295, accessed the global Internet through the main Iranian AS (ITC), as shown on Figure n°10. The architecture of the Afghanistan domestic network is the archetype of architectures where the failure of an AS or the blocking of Iran could lead to a complete blackout of the Afghan Internet. Thus, Iran plays a fundamental role in granting its economic partners an access to the global Internet and can use this access as a tool of influence[26].

## IV) Conclusion

This paper offers new perspectives on the use of BGP, routing policies and network architectures to understand the geography and geopolitics of cyberspace. We used an in-depth analysis of the Iran domestic network and its connectivity to the global Internet as a case study illustrating the use of the geopolitical approach to understand the strategic dimension of technical matters. Through the mere observation of BGP architecture and incidents, we identified several ways in which Iran could have deliberately used cyberspace to achieve its strategic goals of asserting its own power both domestically and more widely in the Middle East. Although technical observations alone cannot tell whether this resulted from a coordinated strategy from the regime, or not, it shows interesting features of the Iranian cyberspace and provides the regime with strategic assets, as well as vulnerabilities.

We demonstrated that Iran's BGP structure is organized around ASes controlled by the government that connect the domestic network to the global Internet, providing the government with an Internet "kill-switch" to fully disconnect Iran from the global Internet. By limiting the number of ISPs and ASes directly connected to the outside network, Iran has created a frontier between its domestic network and the rest of the Internet. However, at the same time Iran has built a lively ecosystem of ASes with a rich set of paths within the country. The Iranian domestic network is therefore very resilient within its borders, due to its low level of centralization and high complexity. We compared the evolution of the Iranian network with other countries in the Middle-East and observed that while Iran is not the only country that have increased its control over its domestic network, it has succeeded in increasing its internal complexity along with increasing control. This later point distinguishes Iran from other countries in the Middle-East and demonstrates the existence of a strategy of control through the BGP architecture. We thereafter observed and evaluated how Iran has leveraged this domestic network structure both to implement an active strategy of censorship based on BGP hijacks and outages, and to attract traffic from their neighbors and emerge as a major connectivity provider in the middle east.

---

[25] BGPHurricane Electric: https://bgp.he.net/AS39216#_graph4 et https://bgp.he.net/AS344217#_graph4 Accessed August 14, 2019.

[26] BGPHurricane Electric: https://bgp.he.net/AS39216#_graph4 ; https://bgp.he.net/AS344217#_graph4 Accessed August 14, 2019.

What is notable in our finding is that we used solely BGP and routing information in our analysis, illustrating how these tools are relevant to develop a geopolitical analysis of the cyberspace. The present methodology that was focused on Iran cyberstrategy, could be extended to other geographical contexts. Our methodology can also be enriched by other classical sources of information, like online content. This can lead to a more comprehensive understanding of the geography of cyberspace and the cyberstrategies of Internet actors. As the environment is highly dynamic, longitudinal observations could provide an interesting window into the evolution of state's strategies according to the geopolitical context and how they contribute to shape cyberspace.